\documentclass[a4paper,preprint]{revtex4}
\usepackage{graphicx}
\usepackage{courier}
\usepackage{amsmath}
\usepackage{prettyref}

\newcommand{\MgHion}{MgH$^+$(X$^1\Sigma^+$)}
\newcommand{\XMgH}{MgH(X$^2\Sigma^+$)}
\newcommand{\AMgH}{MgH(A$^2\Pi$)}
\newcommand{\Rbground}{Rb($^2$S)}
\newcommand{\Rbion}{Rb$^+$($^1$S)}
\begin{document}

\title{Molecular ions in ultracold atomic gases: computed electronic interactions for
  \MgHion\, with Rb}

\author{Mario Tacconi}
\author{Franco A. Gianturco}
\thanks{Corresponding author: Dep. of Chemistry, University of Rome {}``La
Sapienza'', P. A. Moro 5, 00185, Rome, Italy. Fax: +39-06-49913305. }
\email{fa.gianturco@caspur.it}
\affiliation{Department of Chemistry and CNISM,
University of Rome La Sapienza, Piazzale A. Moro 5, 00185 Rome,
Italy}

\begin{abstract}
The electronic structures of the manifold of potential energy surfaces generated in the lower energy range by the
interaction of the \MgHion\, cationic molecule with \Rbground\, neutral atom are obtained over a broad range of
Jacobi coordinates from strongly correlated \emph{ab initio} calculations which use a Multireference (MR) wavefunction within
a Complete Active Space (CAS) approach. The relative features of the lowest five surfaces are analyzed in terms of
possible collisional outcomes when employed to model
the ultracold dynamics of ionic molecular partners.
\end{abstract}

\maketitle

\section{Introduction}
The experimental and computational advances in our capacity of producing and trapping samples of cold (below 1 K) or ultracold
(below 1 mK) molecules in the gas phase have made tremendous progress in recent years and have dramatically enhanced our
detailed understanding of the processes at hand within their nanoscopic evaluation~\cite{fake:ref, exp:doyle_1998nat, julienne_2006rmp, theo:bodo_2006irpc}. The ensuing cold and ultracold molecular
species have many possible applications such as allowing for the accurate measurement of fundamental physical properties~\cite{fake:ref}, the possible evaluations
of small energy differences between enantiomers~\cite{stohner_2005ch} and the observation of temporal changes of the fine-structure constant~\cite{erhudson_2006prl}. Furthermore, dipolar
species have been suggested as qubits in quantum computers~\cite{demille_2002prl} and polar gases in general, neutral and ionic, are expected to exhibit even more
marked novel features due to their environment~\cite{baranov_2002psr}.
In addition, translationally cold molecular ions, and polar ions at that, can also become embedded inside Coulomb crystals~\cite{schiller_2006jpb} and provide ideal targets
for a large variety of investigations like high precision spectroscopic measurements and state selected partners in ionic reactions~\cite{vogelius_2006jpb}.
Earlier suggestions in this experimental area include the analyses of the Ca$^+$/Na system by the Storrs group~\cite{exp:makarov_2003pra}, while the choice of a molecular partner as we shall be examining here is currently being considered by various experimental groups as a possible option~\cite{schiller_privat}.
The insertion of a molecular ion partner
within a Coulomb crystal~\cite{raizen_1992pr}, i.e. their phase transition to an ordered state at ultralow temperatures in the range of a few millikelvins, is one
of the intriguing steps for the subsequent handling of such ions within a broad variety of molecular processes.
The present study therefore intends to provide, within a fully ab initio formulation of the problem, an accurate description of the interaction forces which
drive some of the possible collisional processes at such low temperatures. In particular, we wish to present the energy details and the spatial features
of the various potential energy surfaces (PES) which become relevant in the direct process that involves \MgHion\, in collision with \Rbground , both partners being current candidates in cold trap experiments~\cite{schiller_privat}. Because
of the differences between the ionization potentials of the two partners we shall also show that the \XMgH\, molecule interacting with \Rbion\, is indeed
an important outgoing channel, which provides an interesting \emph{reactive} outcome of the process and which remains open even at ultralow energies.
The Section~\ref{sec:compdet} therefore discusses our computational details while Section~\ref{sec:result} presents our results and analyses the novel features of our
final PES's. Conclusions are given by the last Section~\ref{sec:concl}.

\section{A computational scheme for the interaction forces}\label{sec:compdet}
\subsection{The RbMgH$^+$ electronic structure}
\begin{figure}[floatfix]
 \includegraphics[width=1.0\textwidth]{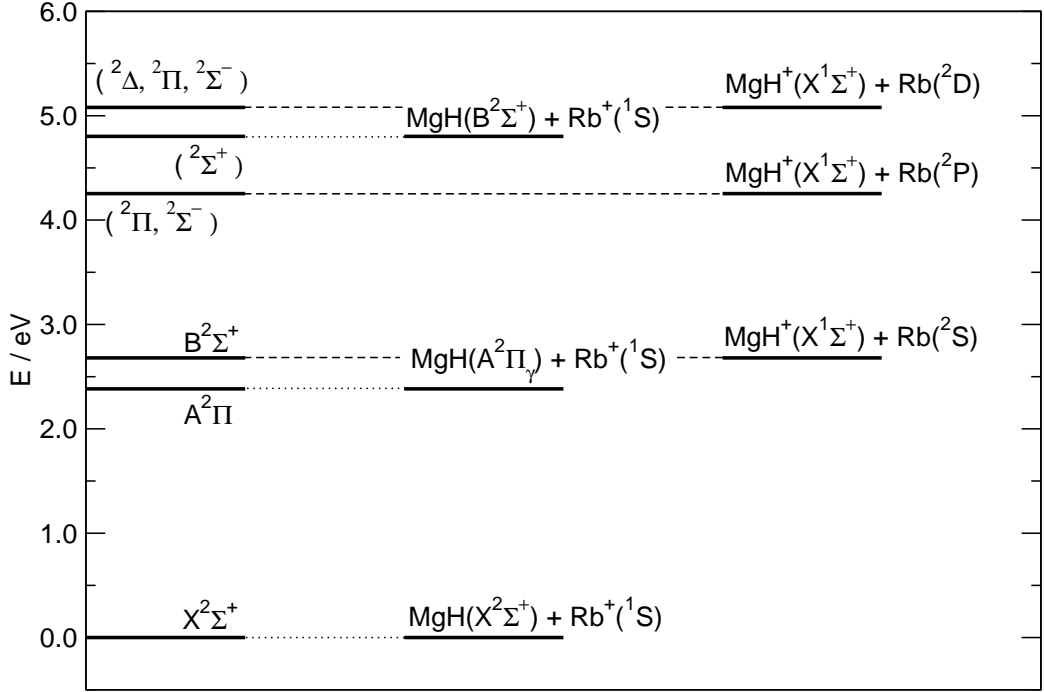}
 \caption{Relative energetics of the atomic and molecular asymptotic states, with their
           resulting short-range couplings terms given for the linear configurations, of the the MgH interacting with a Rubidium ion and
      for the MgH$^+$ interacting with a Rubidium atom.
          See text for details.} \label{fig:rbmgh_levels}
\end{figure}

\begin{figure}
 \includegraphics[width=.30\textwidth]{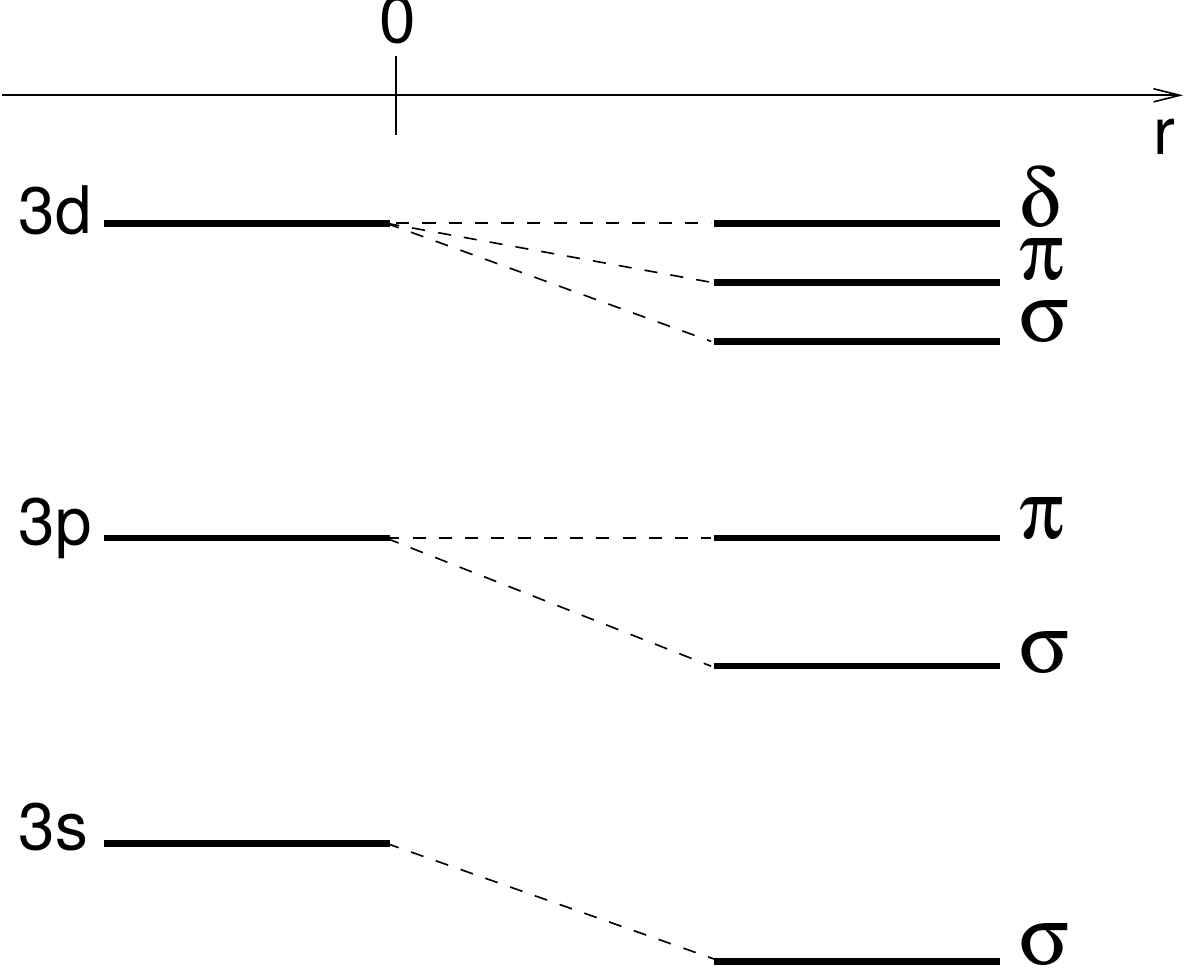}
 \caption{United atoms correlation diagram for the molecular orbitals of the MgH molecule.
   The united atom limit of MgH is the Al atom which has the electronic configuration: [Ne]$3s^23p^1$.}\label{fig:mgh_uamo}
\end{figure}

In order to become familiar with the terms and the physical  features of the chosen partners in the iontrap, it is useful to remind ourselves of the way in which
their electronic features are put together from an ab initio viewpoint.
A direct comparison between the Ionization Potentials (IP) of the MgH(X$^2\Sigma^+$) and Rb($^2$S) reveals that the ground state of the complex molecular
ion RbMgH$^+$ has its lowest energy correlation  with the MgH(X$^2\Sigma^+$) + Rb$^+$($^1$S) species.
Consequently, the electronic state of the ion complex that correlates with MgH$^+$(X$^1\Sigma^+$) + Rb($^2$S), i.e. with fragments which do not undergo charge-exchange, is an excited state from the RbMgH$^+$
electronic manifold. If we start our analysis with the simpler case of a collinear geometry, an appropriate generalization of the Wigner-Witmer correlation rules for diatomic molecules (see for example pagg. 281-284 of~\cite{theo:herzbergIII}) allows us to derive the electronic state manifold of the complex using only the knowledge of the spatial and spin symmetry properties of the electronic states of the combining fragments. It is worth noting here that the use of the collinear geometry does not limit the generality of our results: it is always
possible to univocally resolve the electronic states symmetry species for a particular spatial configuration of a molecule into
those of a molecular geometry of lower symmetry by means of group theory. Furthermore, if we combine that qualitative
information with the available spectroscopic data for the MgH and MgH$^+$, and the IP of Rb and MgH, we would obtain the relative
energies for the complex electronic states.
In this particular case, we have taken as origin of the energy axis the MgH(X$^2\Sigma^+$) + Rb$^+$($^1$S) energy at infinite
separation distance. The relative energy of the MgH$^+$(X$^1\Sigma^+$) + Rb($^2$S) asymptote, which is the lowest energy pathway to
dissociate into the MgH$^+$ molecular ion and a neutral Rb atom, with respect to the origin of the energy axis is simply the electronic energy difference between
the MgH(X$^2\Sigma^+$) and the Rb($^2S$) IPs (about 2.69 eV). The asymptotic relative energy of any electronic state of the triatomic
complex which correlates with an excited electronic state of one, or both of the fragments, is simply obtained by adding the electronic
excitation energy to the energy of the appropriate asymptote. For example,  the MgH$^+$(X$^1\Sigma^+$) + Rb($^2$P) energy equals the
energy of the MgH$^+$(X$^1\Sigma^+$) + Rb($^2$S) plus the Rb electronic $^2$S-$^2$P transition energy. In this way it is possible
to determine the relative energy ordering of the triatomic complex electronic states, at least for an infinite separation between
the fragments. In fact, it should be clear that this approach while giving an exact idea of the possible electronic states, says
nothing about their stability and even less about which nuclear conformation is the most stable. Nevertheless, this qualitative
information provides valuable help in the choice of the more suitable \emph{ab initio} method which should be used in the actual electronic
structure calculation.
Our figure~\ref{fig:rbmgh_levels} summarizes the results obtained from this type of analysis. At the very right side of the
figure~\ref{fig:rbmgh_levels} we report the possible dissociation channels in the molecular ion MgH$^+$ and neutral Rb for the energy
range we have considered relevant for our purposes in this study, while in the center the possible dissociation pathways into neutral MgH
and Rb$^+$ are represented. Finally, proceeding from the right to the left of figure~\ref{fig:rbmgh_levels} the Wigner-Witmer correlation
rules bring the fragments electronic states into the complex ones.
A perusal of figure \ref{fig:rbmgh_levels} suggests that a reasonable description of the RbMgH$^+$  manifold of states could be
obtained as a summation of three subsets of electronic states well separated in energy: in the first subset we put the system
ground state, in the second one the A$^2\Pi$ and the B$^2\Sigma^+$ states, while the third one collects a number of very high
energy electronic states which will eventually correlate with the Rubidium excited states and the MgH(B$^2\Sigma^+$) state.
It is also worth noting here that all the electronic states employed in the present scheme have the same spin symmetry: they are
all doublets. With that information in mind, it is clear that an \emph{ab initio} description of the RbMgH$^+$ electronic structure
should be able to treat at least the ground state and the first two excited states in a balanced manner. In other words,  in a
case like the present one the use of a Multireference (MR) electronic wavefunction is mandatory. We have therefore designed a MR
wave function within the State Averaged Complete Active Space (SA-CAS) approach: thus, the problem of the choice of a set of
N-electron wave functions (Slater Determinants or Configuration State Functions) is recast into a partition of  the molecular
orbital space in
\begin{itemize}
  \item
    active orbitals, which can have different occupation numbers (0,1,2) and
  \item
    inactive orbitals, which are doubly occupied in all the configurations.
\end{itemize}
The electrons which populate the active orbitals are then called the active electrons. It is worth noting that, in our calculation the active electrons correspond exactly
to the ones commonly defined in chemistry as the valence electrons. In other words, we choose those molecular orbitals that are supposed to generate within the CAS
procedure all the N-electron configurations which are important for the description of the electronic states we are interested in.
A prior knowledge of the electronic structure of the isolated fragments is therefore helpful in guiding the choice of the active orbital space within the complex:
in figure \ref{fig:mgh_uamo} a united-atom correlation diagram allows us to determine the symmetry and gives us an idea of the relative energies of the valence
molecular orbitals of MgH and MgH$^+$ by considering the corresponding structures of Al and Al$^+$. Following the aufbau principle, we therefore see that the first two low-lying electronic states, X$^2\Sigma^+$ and A$^2\Pi_\gamma$,
could be described by single electronic configurations: $3s\sigma^23p\sigma^1$ and $3s\sigma^23p\pi^1$ respectively. The case of the next higher energy state B$^2\Sigma^+$ is
not so straightforward because more than one configuration could be important to describe it: $3s\sigma^23p\sigma^1$, $3s\sigma^13p\sigma^2$,
$3s\sigma^13p\sigma^13d\sigma^1$, etc. In the case of the MgH$^+$(X$^1\Sigma $) state the single configuration $3s\sigma^2$ could be the one which manages to account for the main
features of that electronic state. We used that qualitative argument for the choice of a suitable orbitals active space for the calculation of the X$^2\Sigma^+$, A$^2\Pi$ and
B$^2\Sigma^+$ MgH electronic states: $3s\sigma$, $3p\sigma$, $3p\pi_x$, $3p\pi_y$, $3d\sigma$ and a $\sigma^*$ anti-bond molecular orbital which has to be included in order
to correctly describe the bond dissociation. This active space produces the CAS(3,6) (3 active electrons in 6 active orbitals) multireference wavefunction that we used
as reference in the Multi Reference Configuration Interaction with Single and Double excitations (MRCI-SD) calculation of the MgH potential energy curves (PECs) reported
in the upper panel of figure~\ref{fig:mgh_pes}. A completely analogous orbital
active space was used to calculate the PECs of MgH$^+$ low lying electronic states (reported in the lower panel of figure~\ref{fig:mgh_pes}), here the only difference is
the number of active electrons: there are  two of them. It is interesting to note that the value of the coefficients of the Configurations Interaction (CI) expansion actually reflects
the simple qualitative argument based on the aufbau principle and the molecular orbital scheme obtained from the united atom correlation diagram. In fact, as we can see
from table~\ref{tab:CIcoeff}, the coefficient of the $3s\sigma^23p\sigma^1$  configuration is about 0.95 for the X$^2\Sigma^+$ state, while the A$^2\Pi$ is chiefly described
by the $3s\sigma^23p\pi^1$ configuration (CI coefficients of 0.96 ). As expected, in the case of the B$^2\Sigma^+$ state more than one electronic configuration is important:
the CI coefficients of the $3s\sigma^13p\sigma^2$ and  $3s\sigma^23d\sigma^1$ configurations have comparable values: 0.77 and 0.55 respectively. It is worth noting that the
configurations in which the $\sigma^*$ is occupied essentially don't contribute to the CI expansion, at least at the  interatomic distances around the equilibrium value.
\begin{table}
  \begin{tabular*}{0.60\textwidth} %
    {@{\extracolsep{\fill}}c||c|c||c||c}
    \hline
    conf.    &  X$^2\Sigma^+$ & B$^2\Sigma^+$ & conf.& A$^2\Pi$\\
    \hline
    2+00 00  &  0.9537  &  -0.1410  & 2000 +0   &  0.9619 \\
    +200 00  &  0.0683  &   0.7480  & ++00 -0   &  0.1253 \\
    20+0 00  & -0.1951  &  -0.5526  & 0200 +0   & -0.1206 \\
    +000 02  &  0.0433  &  -0.1626  & 0+-0 +0   & -0.0762 \\
    +000 20  &  0.0433  &  -0.1626  & 0+0- +0   &  0.0576 \\
    200+ 00  &  0.0204  &  -0.0823  &           &         \\
    +-+0 00  &  0.0792  &  -0.0780  &           &         \\
    0+00 20  & -0.0328  &   0.0662  &           &         \\
    0+00 02  & -0.0328  &   0.0662  &           &         \\
    +-0+ 00  & -0.0003  &   0.0606  &           &         \\
    ++-0 00  & -0.0538  &   0.0146  &           &         \\
    +020 00  & -0.0089  &  -0.0501  &           &         \\
    \hline
    \end{tabular*}
  \caption{Reference coefficients larger than 0.05 of the MgH CAS(3,6) MRCI-SD wavefunction. The columns labeled by ``conf.'' report the occupation number
    of the active orbitals (i.e. the electronic configurations). More precisely in those columns, the first four occupation numbers refer to the $\sigma$-like orbitals
    ($A'$ symmetry): $3s\sigma$, $3p\sigma$, $3d\sigma$ and $\sigma^*$, while the last couple of occupation numbers refers to $\pi$-like molecular orbitals
    ($A'$ and $A''$ space symmetry). The symbol + or - stands for orbitals occupied by only one electron in an $\alpha$ or $\beta$ spin state,
    respectively. The columns labeled by the electronic states term symbols report the coefficients of the CI expansion for the particular electronic state.}\label{tab:CIcoeff}
\end{table}
On the Rb atom side the situation is much simpler: asymptotically we are dealing with two species, Rb$^+$($^1$S) and
the Rb($^2$S), both of which could be described by one electronic configuration: [Kr]$5s^1$ and [Kr]$5s^0$. Using that information one step further,
we chose as active orbitals those complex molecular orbitals which correlate asymptotically with the MgH, MgH$^+$, Rb, Rb$^+$ relevant configurations;
namely, MgH($3s\sigma$, $3p\sigma$, $3p\pi_x$, $3p\pi_y$, $3d\sigma$), Rb($5s$). This orbital space defines the CAS(3,6) wavefunction that we finally used as a
reference space in the MRCI-SD calculations of the RbMgH$^+$ PESs.
\subsection{Details of the PES's calculations}
\begin{figure}
 \includegraphics[width=.90\textwidth]{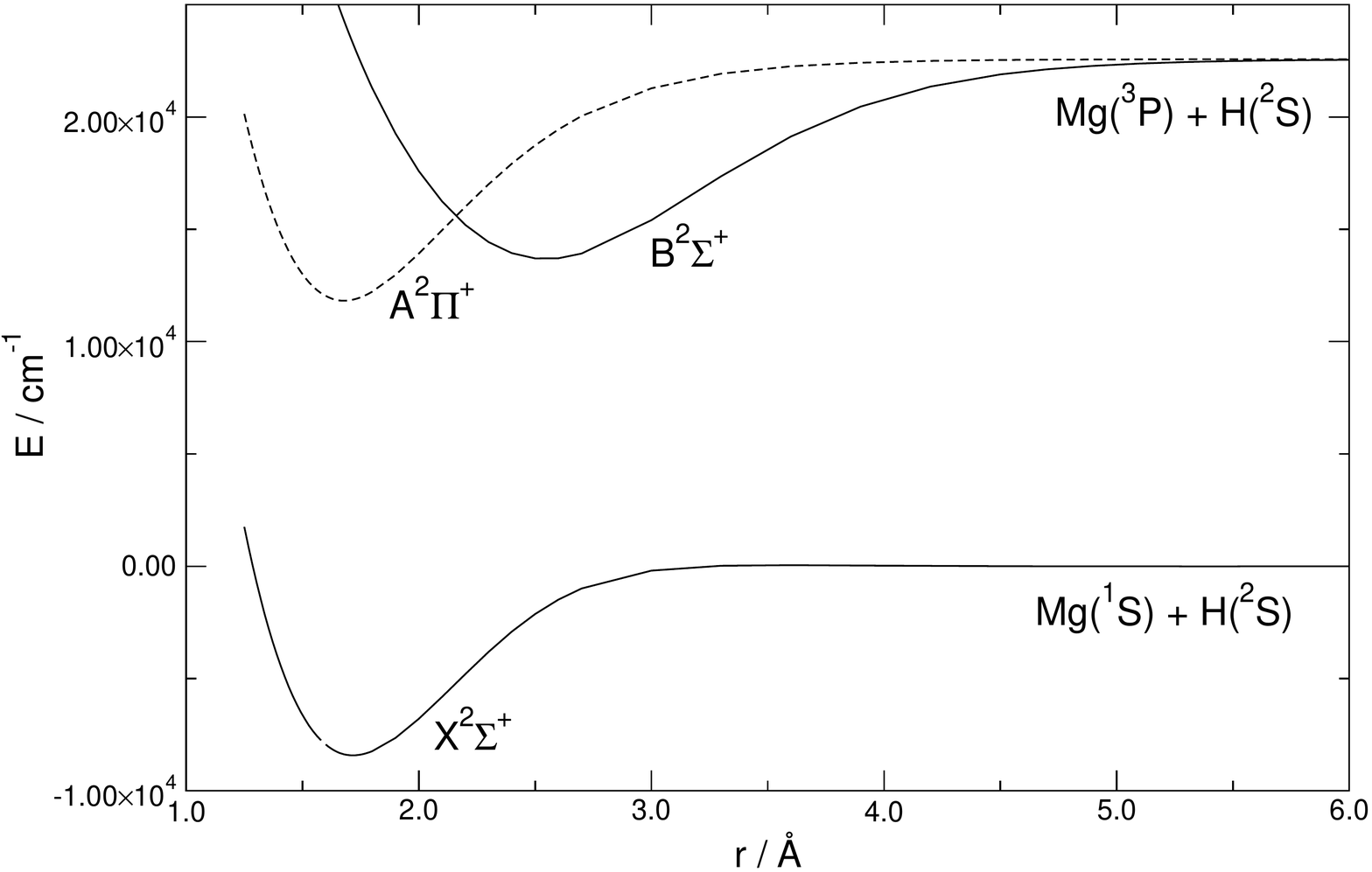}
 \includegraphics[width=.90\textwidth]{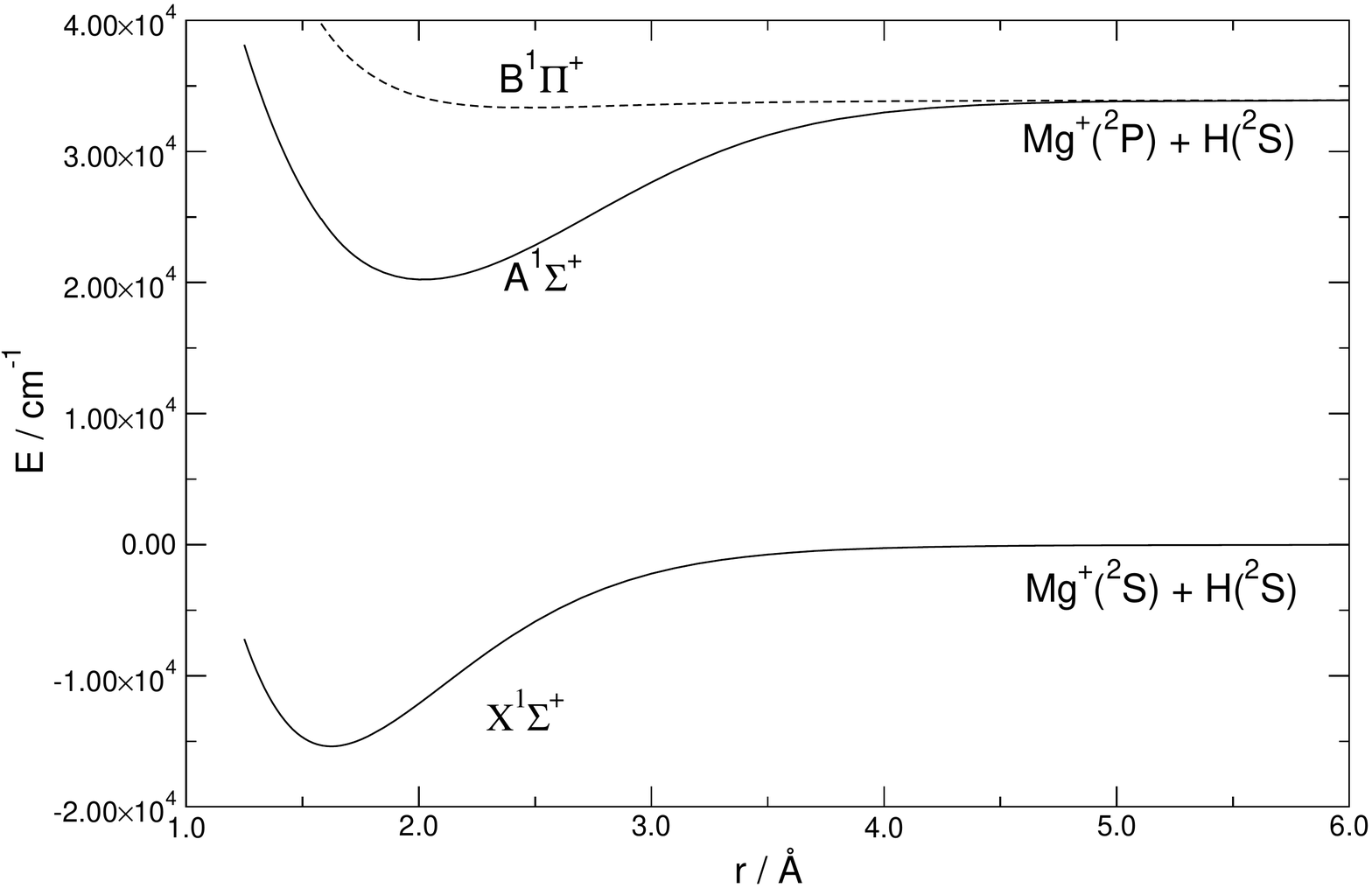}
 \caption{Upper panel: computed ab initio potential energy curves for the lower lying doublet electronic states of the MgH. See main text for
   computational details. \\
   Lower panel: Same as in the upper panel but for the ionic case of MgH$^+$}\label{fig:mgh_pes}
\end{figure}

The geometry of the MgH-Rb ion complex is described in Jacobi (r,R,$\theta$)  coordinates where r is the distance between the atoms within the MgH molecule, R indicates the distance
between the center of mass of MgH and the Rb atom and the angle $\theta$ is taken between R and r: the  $\theta=0^o$ orientation corresponds here to the MgH-Rb collinear configuration with the Rb atom approaching the H side of the neutral/ionic molecule: in the present calculations the distance r  was kept fixed at the calculated
equilibrium distance for the MgH$^+$(X$^1\Sigma^+)$ ($r=r_e=1.6250$ \AA). The equilibrium distance of the MgH$^+$(X$^1\Sigma^+$) electronic states was extrapolated from the
ground state PEC reported in the lower panel of figure~\ref{fig:mgh_pes}. The electronic structure calculations have been carried out on a discrete grid defined by
varying the Jacobi coordinates R and $\theta$. The R coordinate was allowed to vary between 3.00 and 16.00 \AA. A radial grid step size $\Delta R$ of 0.25 \AA\,
was used in the interval [3.00, 7.50] \AA, in the interval [7.50,10.0] \AA\, $\Delta R=0.50$, while in the interval [10.0,16.0]\AA\, $\Delta R$ was 1.00 \AA. We defined an
angular grid in the interval [0$^o$,180$^o$] with a step size $\Delta\theta=15^o$. The total number of grid points and, eventually, the number of computed ab initio
energies was 450. The CAS(3,6) wavefunction outlined in the previous section was used, for each grid point, in a State-Averaged MCSCF (SA-MCSCF)
calculation of the lowest 4 doublet states of the RbMgH$^+$ (the three $\Sigma^+$ and the one $\Pi$). However, because the calculations were carried out in $C_s$ symmetry,
the actual number of averaged states is indeed 5 (four $A'$ and one $A''$) because the $\Pi$ state splits into two components ($A'$, $A''$) which are fully degenerate
only at collinear geometries. The SA-MCSCF therefore generates a common set of orbitals for the 5 states which  in turn we have used for the subsequent MRCI-SD calculations
which we expect to correctly account for electron correlation effects: all the electrons, except the Mg (1s$^2$), were treated as correlated in the present calculations.
It is worth noting that the
highest $^2\Sigma^+$ state was included only to ensure correct convergence of the SA-MCSCF wavefunction at all the considered
geometries and doesn't really have a simple physical meaning: from figure \ref{fig:rbmgh_levels} one can see that the actual electronic states above the B$^2\Sigma^+$ are
those which correlate with MgH$^+$(X$^1\Sigma^+$)+Rb($^2$P), i.e. with our present asymptotic state. Those electronic states, however,  cannot be described by the CAS(3,6) wavefunction because the active orbitals
space does not include molecular orbitals of the complex which correlate with the Rb 5p atomic orbitals. However, this is not a real problem in the present work since
we would like to gather information on the possible relaxation processes that specifically involve MgH$^+$(X$^1\Sigma^+$) interacting with Rb($^2$S). In other words,
we can disregard what is energetically above the B$^2\Sigma^+$ state of the complex and focus only on the first three lower lying electronic states of the latter.
The gaussian basis set we selected was the cc-pVQZ(spdf) (16s,12p,3d,2f) contracted as [6s,5p,3d,2f] for the Mg~\cite{basis:NaMgpVQZ}, the cc-pVQZ(spd) (6s,3p,2d) contracted as [4s,3p,2d]~\cite{basis:HpVQZ}. The Rb atom was treated within the Effective Core Potential (ECP) framework in order to reduce the number of electrons  treated explicitly but to still account for the relativistic effects
on the core electrons. The ECP used was the ECP28MDF from the Stuttgart Group~\cite{comp:ecp_rbmdf}; this ECP leaves the $4s^24p^65s^1$ Rb electrons explicitly described while the effect of the others
is included within a  pseudopotential formulation. The gaussian basis set chosen for the Rb atom was the companion basis set published along with the ECP28MDF parameters: the basis
is (13s,10p,5d,3f) contracted as [7s,6p,5d,3f]~\cite{comp:ecp_rbmdf}. All the calculations have been done by using the MOLPRO 2000.6 suite of programs~\cite{comp:molpro_cas, comp:molpro_ci}.
In order to obtain the interaction energy from the calculated electronic energy, we used as ``zero of the interaction energy'' the electronic energy of the
system calculated at  R=300 \AA. At this distance the angular variation of the ground state energy is about $10^{-8}\,E_h$ (about 0.002 $cm^{-1}$). The remaining
interaction energy at this distance could be estimated as not greater than $0.04\,cm^{-1}$ which is the difference between the ground state energy calculated at R=300.0\AA\
and the one calculated at R=275.0\AA.
\begin{table}
  \begin{tabular*}{0.85\textwidth} %
    {@{\extracolsep{\fill}}c|c}
    \hline
    basis & I.P. (eV) \\
    \hline
    cc-pVTZ(sp)+ECP+CPP & 6.627\\
    cc-pVQZ(sp)+ECP+CPP & 6.815\\
    Full Electron Mg(spd) H(sp) cc-pVTZ & 6.857\\
    Full Electron Mg(spdf) H(spd) cc-pVQZ & 6.862\\
    \hline
    \end{tabular*}
  \caption{Computed Ionization Potential (IP) for MgH using different basis sets with or without an ECP and the Core Polarization Potential (CPP)
    for the Mg atom. All the calculations have been done using a RCCSD-T (Coupled Cluster) wave function.
    The experimental MgH and MgH$^+$ equilibrium distances were used. The IP values are not corrected for the Zero Point Energy (ZPE)
    within the molecular hydride.}\label{tab:IP}
\end{table}
\begin{table}
  \begin{tabular*}{0.85\textwidth} %
    {@{\extracolsep{\fill}}c|c|c||c}
    \hline
    method & $^2$P (cm$^{-1}$)& $^2$D (cm$^{-1}$) & I.P.(eV)\\
    \hline
    MRCI-SD & 12731.4 & 19605.2 & 4.147 \\
    RS2 & 12849.4  & 20367.4  & 4.167 \\
    RS3 & 12647.2  & 20187.0  & 4.132 \\
    exp. & 12697.75 & 19355.43 & 4.177 \\
    \hline
    \end{tabular*}
  \caption{Computed Rb excited energy levels and ionization potential using the ECP28MDF and different \emph{ab initio} methods. The experimental
    values (averaged over the spin-orbit splitting) are also reported for a comparison. The excited energy levels are
    all in cm$^{-1}$ while the calculated IP are in eV. The RSn reported in the table, with n=2,3, are roughly equivalent to the CASPT2 and CASPT3
    perturbative methods respectively.}\label{tab:RbExt}
\end{table}
Because of the important role played by the ionization potential of the molecular partner in the present study, we report in Table~\ref{tab:IP} the values obtained from our test calculations for the MgH molecule
and in Table~\ref{tab:RbExt} the calculated IP along with the lower-lying energy levels of the Rb atom. The molecular data although not corrected for the Zero Point Energy (ZPE) effect, employ the MgH
equilibrium geometry of 1.77 \AA\, (from ref.~\cite{saxon_1978}) and the MgH$^+$ equilibrium bond length of about 1.64 \AA\, (from ref.~\cite{olson_1979}). Since our calculated values turned out to be very close to the experiments (see before) we see in the molecular case that the cc-pVQZ basis set choice converges to a stabilized estimate of
the IP value. The atomic calculations also appear to get very close to experiments and surmise a difference between the two asymptotic charge-exchange systems of about $2.2 \cdot 10^4\,cm^{-1}$. To make pictorially clearer the relative shapes of the
 molecular potential curves which are relevant for the present study, we are presenting in the two panels of figure~\ref{fig:mgh_pes} the low-lying electronic states of MgH (upper panel) and the ones related to the molecular ion in the
lower panel. The data clearly show the small geometry change that takes place upon the charge exchange process when we start from the v=0 vibrational state of \MgHion .
\section{Present results}\label{sec:result}
\subsection{The shape of potentials at fixed orientations}\label{sec:resultA}
If we were to adopt for simplicity a purely electrostatic model, which is increasingly valid as the R distance becomes greater, we would recognize that for the
RbMgH$^+$ complex we have two kinds of asymptotic behavior: the complex  can dissociate as  neutral MgH and Rubidium ion  or it can generate an ionic MgH molecule
and a neutral Rubidium atom. It is obvious at this point that the long range behavior of the potential energy function in these two cases is likely to be quite different. MgH is a polar molecule
and therefore  in the first case the long-range potential is dominated by the charge-dipole interaction which is proportional to $\cos(\theta)/R^2$.
Because the MgH dipole orientation, in our case this interaction is attractive for $\theta$ between 0$^o$ and 90$^o$ and repulsive for $\theta$ between 90$^o$ and 180$^o$ degrees. In the second option
the dominant long range term is due to the charge-induced dipole interaction which is proportional to $1/R^4$ and is always attractive since it depends on the spherical dipole polarizability of the neutral Rb partner.
\begin{figure}
 \includegraphics[width=.85\textwidth]{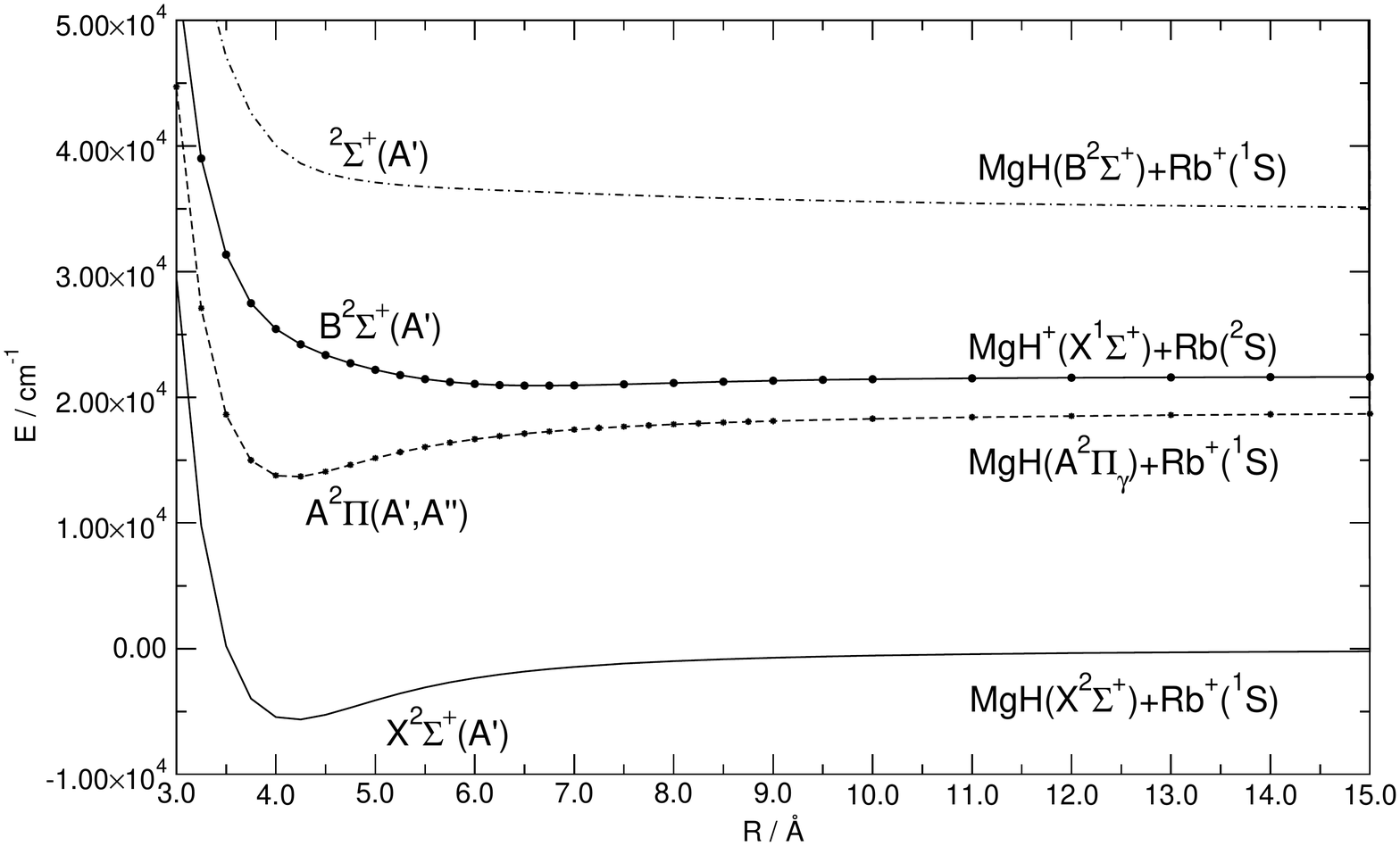}
 \caption{Computed potential energy curves for several asymptotic states for the Rb atom approaching at the H-side of the target molecule
 ($\theta=0^o$). See main text for discussion.}\label{fig:rbmgh_pes_cut0}
\end{figure}
The results reported by figure~\ref{fig:rbmgh_pes_cut0} further underline the characteristic behavior of the relevant surfaces: the lowest curve for the Rb$^+$ approach on the H atom of
the neutral molecular partner clearly shows an attractive well of about $5.6 \cdot 10^3\,cm^{-1}$, placed well below the incoming channel associated with the charge exchange process (the B$^2\Sigma^+$ state) and
also below the more stable MgH$^*$-Rb$^+$ complex of $\,^2\Pi$ symmetry. In other words, the  marked difference in energy between the ionization potential (see Table~\ref{tab:IP} and \ref{tab:RbExt})
generates  the dramatic energy gain that occurs when the interaction is included and which produces the neutral molecule in either an excited or its ground electronic state. The configurations which describe the
approach at the Mg side are reported in figure~\ref{fig:rbmgh_pes_cut180}.
\begin{figure}
 \includegraphics[width=.85\textwidth]{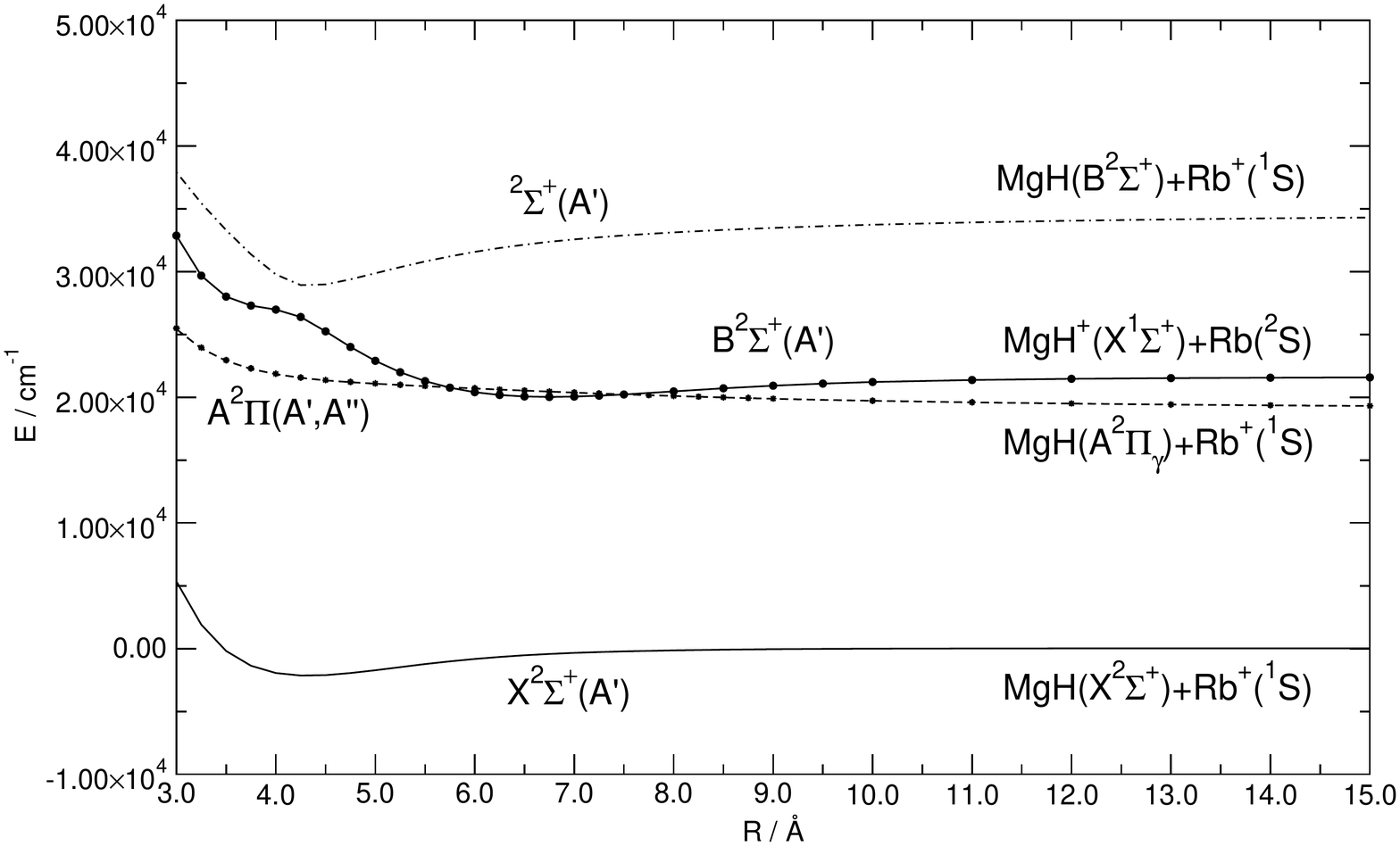}
 \caption{Same potential curves as those given by figure~\ref{fig:rbmgh_pes_cut0}, but this time for the Rb approach on the Mg-side of the
   molecule ($\theta=180^o$).}\label{fig:rbmgh_pes_cut180}
\end{figure}
The basic process remains the same in terms of providing the \XMgH\, as the lower energy molecular partner in the complex, while we see that, for the linear geometries, the possible
splitting of the $\Pi$ state associated with the nearest charge-exchange channel into its two components ($A'$ and $A''$) strengthens the nonadiabatic coupling with the incoming channel since the latter also shows an  $A'$ symmetry component at those
geometries. In other words, the formation of MgH$^+$-Rb complex at large distances (around 6.0\AA) is very strongly coupled with the quasi-degenerate charge-exchange process
into \AMgH\,+\Rbion, which can in turn radiatively decay into the $A'$ complex with \XMgH\, as a final partner molecule.
Thus, while the $\theta=0^o$ orientation indicates a more strongly bound (MgH$^*$-Rb$^+$) complex which could then radiatively decay to its stable ground state but which had
to be reached by a nonadiabatic coupling with a sizable energy gap (see figure~\ref{fig:rbmgh_pes_cut0}), the approach on the other side of the molecule (figure~\ref{fig:rbmgh_pes_cut180}) suggests a much
smaller energy gap for the nonadiabatic event and the more likely radiative emission from a quasi-bound (MgH$^*$-Rb$^+$) complex.
The intermediate situation described by figure~\ref{fig:rbmgh_pes_cut90} report the cuts associated with the $\theta=90^o$ orientation. The $\Pi$ state is now split by the
\begin{figure}
 \includegraphics[width=.85\textwidth]{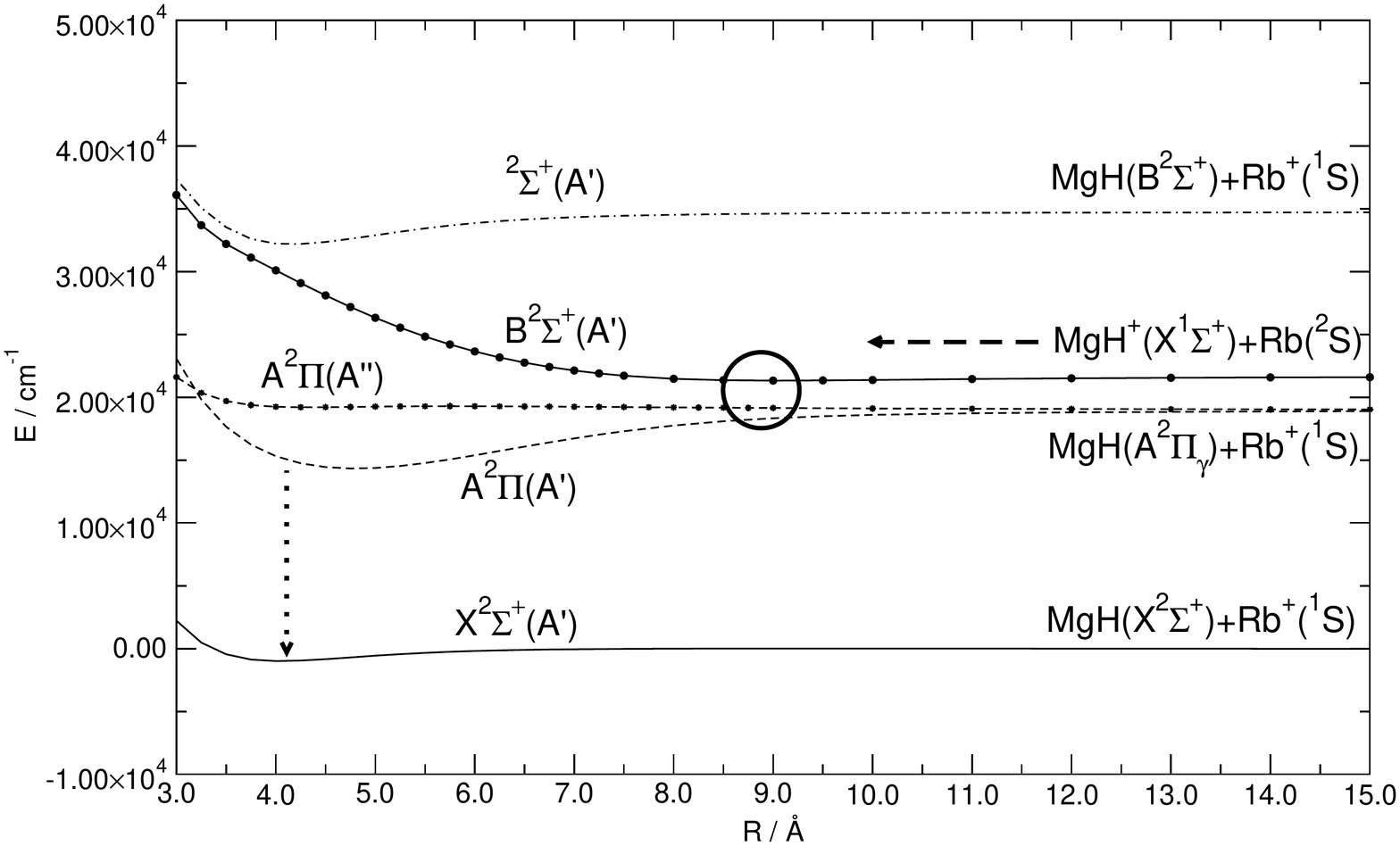}
 \caption{Same computed potential energy curves as in the previous figures~\ref{fig:rbmgh_pes_cut0} and~\ref{fig:rbmgh_pes_cut180} but for the $\theta=90^o$ choice
   of the Jacobi angle. See main text for the discussion of the added symbols onto the computed curves.}\label{fig:rbmgh_pes_cut90}
\end{figure}
interaction and therefore one of its components (the $A'$ state) corresponds to a strongly bound (MgH$^*$+Rb$^+$) charge exchange complex close in energy to the incoming
channel. The latter, bound intermediate complex can now decay to a nearby lower state and form the most stable charge-exchange arrangement. Here again, therefore we see that the analysis of the slopes and feature of the PES already indicates (at least preliminarily) that
the formation of ultracold MgH could occur via charge-exchange complex into an intermediate, bound complex where the molecular partner is in its excited electronic state,
the \AMgH($A'$) state, from which it can further decay by radiative transitions. From the above discussion, therefore, it is clear that the strength of the
relative interactions, and the possible existence of complex bound states within any of the relevant PESs for the title system, are crucial elements for selecting possible pathways to the decay, or excluding it, for the initial system containing the molecular cation. We shall therefore present below the overall fitting of all the relevant interactions which we have discussed above along specific angular cuts and describe their overall feasibility for collisionally coupling radiative states into the final charge-exchange products.

\subsection{The numerical fit of the PESs}\label{sec:resultB}
The calculation of the potential energy surfaces have been carried out, as mentioned before, by considering the initial molecular ion at its equilibrium geometry so that only
a 2-dimensional mapping of the PES's was generated using the previously discussed ab initio procedure.
Each PES was then represented by using the two-center expansion already employed by us in our previous work~\cite{comp:tacconi_2007pra} and described there in greater detail. Here, suffice it to say
that we can represent each adiabatic (Born-Oppenheimer) PES as given by:
\begin{equation}
  V^x(R_a,\theta_a,R_b,\theta_b,R,\theta|r_{eq})=V^x_{sr}(R_a,\theta_a, R_b, \theta_b) + V^x_{lr}(R,\theta)\label{eq:fitf}
\end{equation}
where $sr$ and $lr$ stand for ``short-range'' and ``long-range'' respectively and the labels $a$ and $b$ refer to either atom in the molecular partner.
\begin{equation}
  V^x_{sr}(R_a,\theta_a,R_b,\theta_b) = V^{x(a)}_{sr}(R_a,\theta_a) + V^{x(b)}_{sr}(R_b,\theta_b)\label{eq:sr}
\end{equation}
The two terms in equation~\ref{eq:sr} contain a set of parameters describing the relevant radial coefficients centered on either of the molecular atoms times specific Legendre polynomials which are able to
produce the angular dependence from that atom within the anisotropic interaction. Their actual forms have been given already in our previous work~\cite{comp:tacconi_2007pra, comp:marinetti_2006jctc, comp:tacconi_2007tca}: here, it is obvious that the functional form of the long range part, $V^x_{lr}$, strictly depends on which fragment hosts the positive charge. Two cases can be distinguished:
\begin{itemize}
  \item
    the positive charge is carried by the diatom,
  \item
    the positive charge is carried by the impinging atom.
\end{itemize}
The two options lead to two possible long range behaviors:
\begin{equation}
  V^x_{lr} = C_4\frac{f_4(\beta R)}{R^4} + C_5\frac{f_5(\beta R)}{R^5}\cos\theta\label{eq:lr1}
\end{equation}
in the case of the MgH$^+$ + Rb asymptote, and
\begin{equation}
  V^x_{lr} = C_2\frac{f_2(\beta R)}{R^2}\cos\theta + C_4\frac{f_4(\beta R)}{R^4} + C_5\frac{f_5(\beta R)}{R^5}\cos\theta\label{eq:lr2}
\end{equation}
if the complex dissociates as MgH$^*$ + Rb$^+$. This choice of the long range parts follows directly from the electrostatic model proposed and discussed in section~\ref{sec:resultA}. In both cases as described
by the equations~(\ref{eq:lr1}) and~(\ref{eq:lr2}), the $C_5$ term was included to take into account the dipole-induced dipole interaction effect. All the employed long-range
coefficients are not known and were obtained here from the numerical (linear) fitting of the long range part ($R>8.0$ \AA) of the computed PESs. The computed values of the
long range expansion coefficients are reported in table~\ref{tab:lrcoeff}.

\begin{figure}
 \includegraphics[width=.90\textwidth]{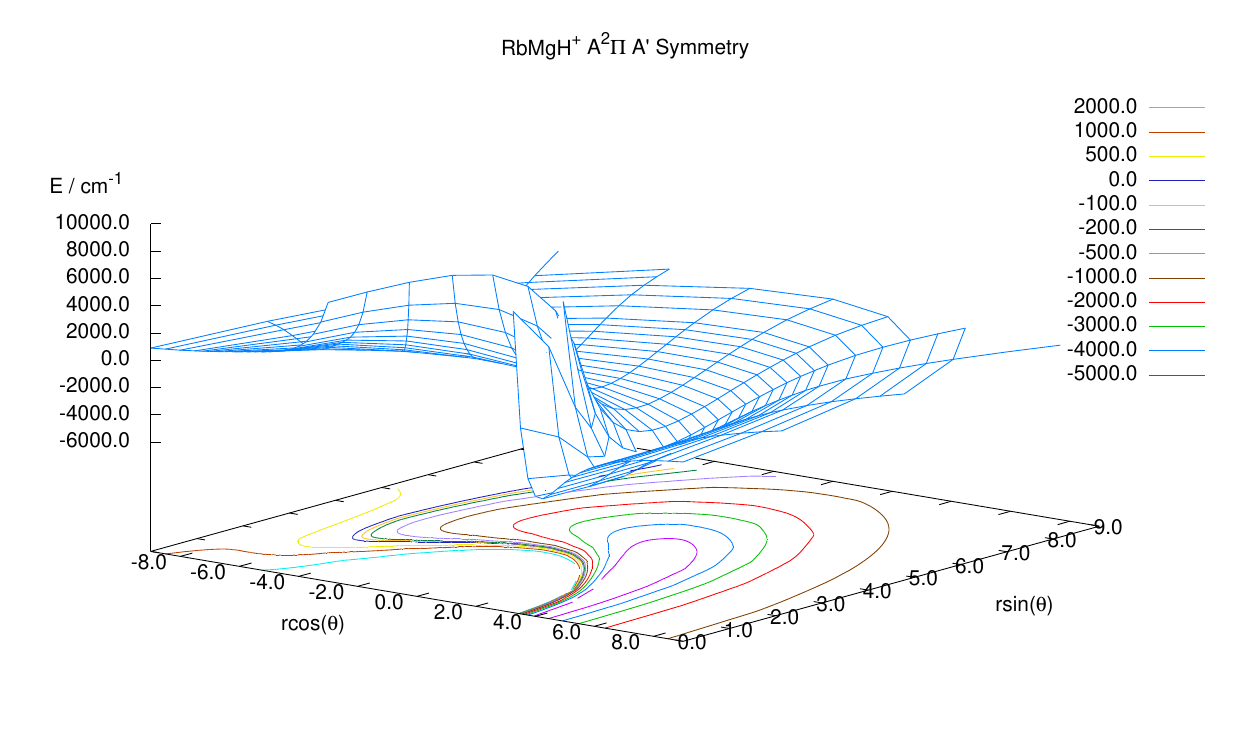}
 \includegraphics[width=.90\textwidth]{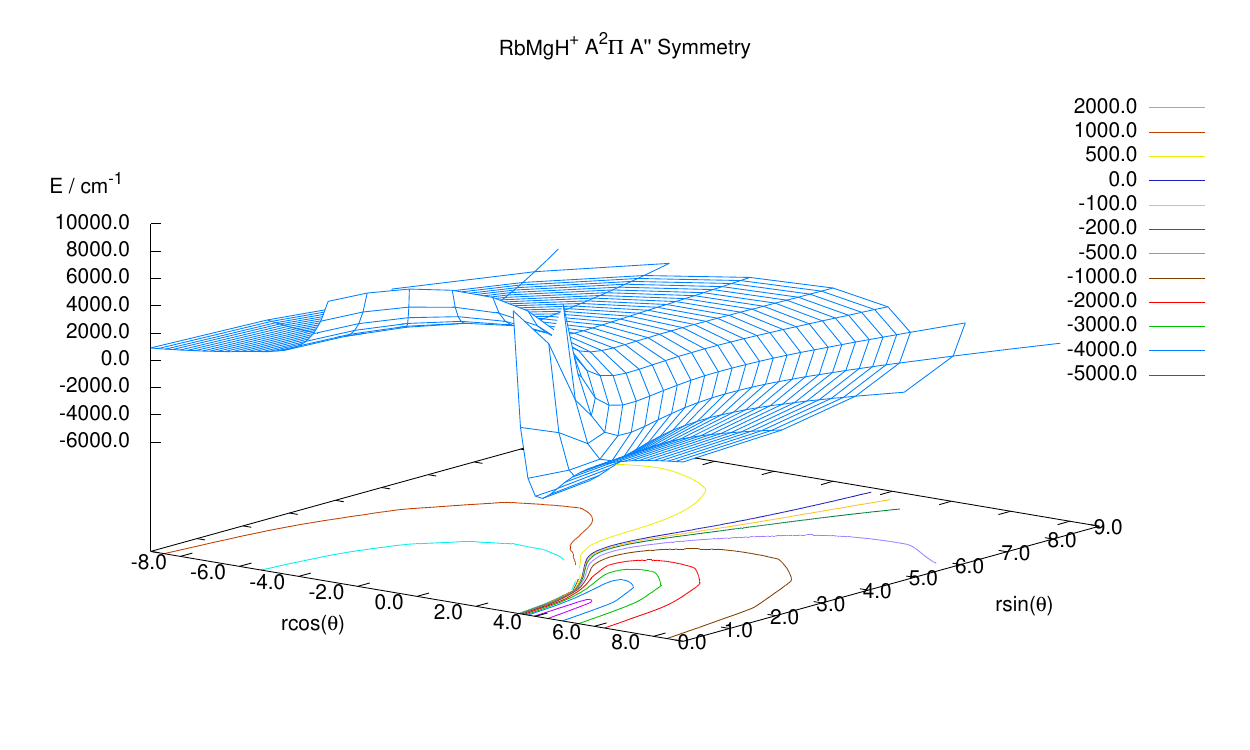}
 \caption{Computed potential energy surfaces for different symmetries. Upper panel: $A'$ symmetry. Lower panel: $A''$ symmetry.
   The two electronic states are degenerate for linear nuclear configuration and give rise to the A$^2\Pi$ term symbol}\label{fig:rbmgh_pes_e2}
\end{figure}

\begin{figure}
 \includegraphics[width=.90\textwidth]{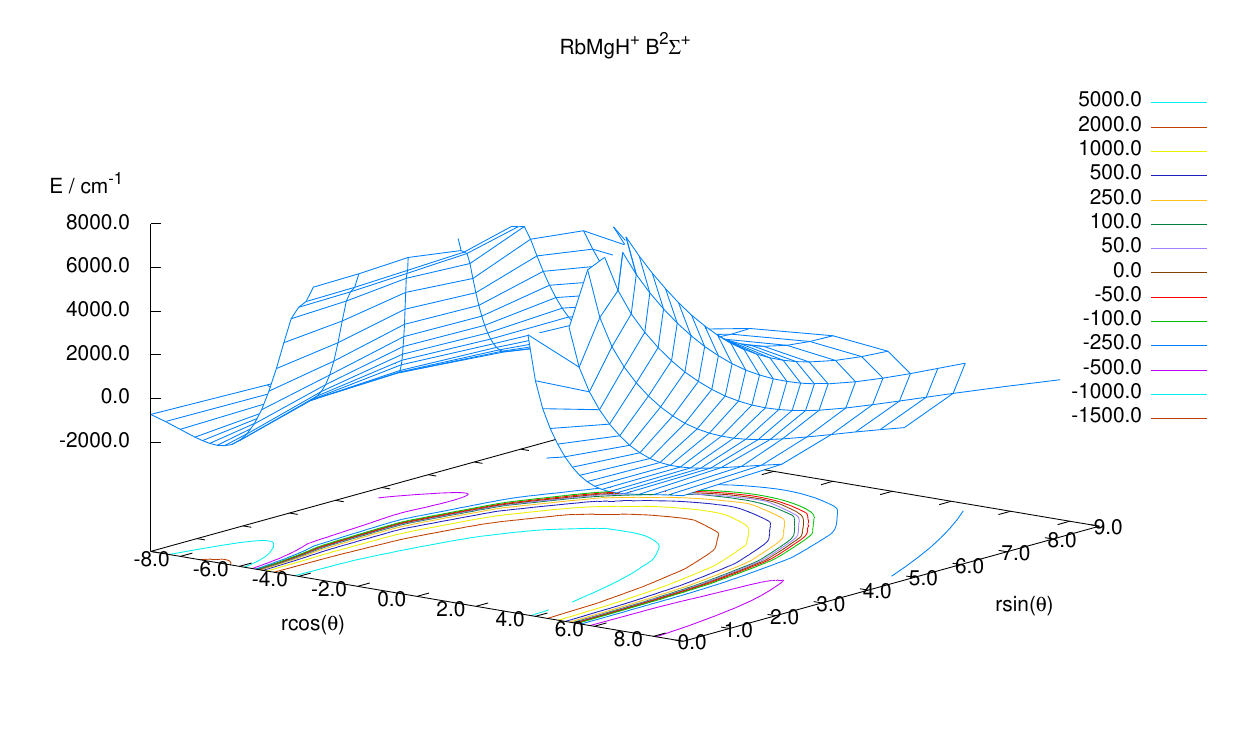}
 \caption{Spatial representation of the potential energy surface of $A'$ symmetry (linear nuclear configuration term symbol: B$^2\Sigma^+$)}\label{fig:rbmgh_pes_e3}
\end{figure}

\begin{table}
  \begin{tabular*}{0.85\textwidth} %
    {@{\extracolsep{\fill}}c||c|c|c}
                  & C$_2$ & C$_4$ & C$_5$ \\
    \hline
    X$^2\Sigma^+$ & $4.2914 \cdot 10^2$  & $-2.9359 \cdot 10^6$ & $9.7780 \cdot 10^6$ \\
    A$^2\Pi$      & $-7.1798 \cdot 10^4$ & $-1.7286 \cdot 10^6$ & $1.9239 \cdot 10^6$ \\
    B$^2\Sigma^+$ &        --            & $-2.9358 \cdot 10^6$ & $1.0603 \cdot 10^7$ \\
    \hline
  \end{tabular*}
  \caption{Computed long range expansion coefficients in $cm^{-1}$). These coefficients have been obtained here from the numerical (linear) fitting of the
    long range part ($R>8.0$ \AA) of the computed PESs.}\label{tab:lrcoeff}
\end{table}
\begin{figure}[floatfix]
 \includegraphics[width=.90\textwidth]{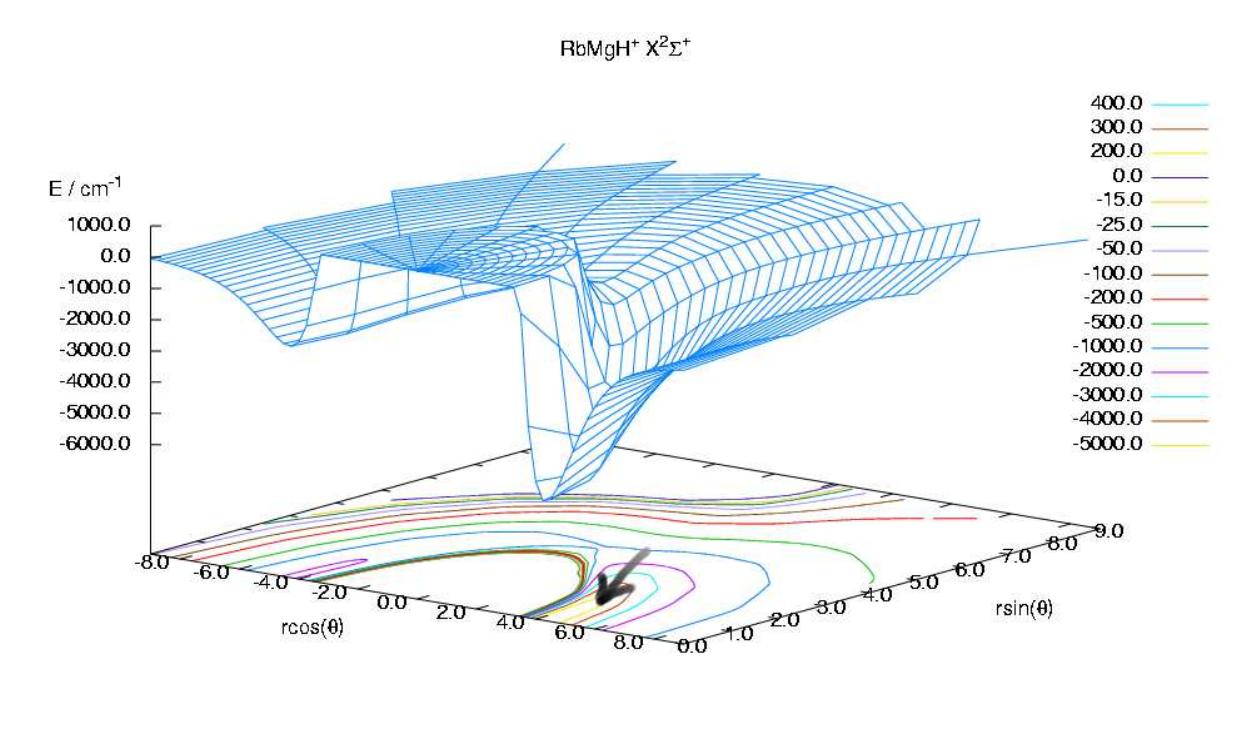}
 \caption{Spatial representation of $A'$ potential energy surface for RbMgH$^+$ (linear symmetry term symbol: X$^2\Sigma^+$).
 The arrow marks the deep well region of the linear configuration.}\label{fig:rbmgh_pes_e1}
\end{figure}
The fitting parameters and the actual analytic representation of the present PESs are available on request from the authors.
A pictorial view of the lower electronic state is given by the 3D representation of figure~\ref{fig:rbmgh_pes_e2}, where the deep attractive well in the collinear geometry with Rb approaching the
H-side of the molecule along charge-exchange surface, as presented by figure~\ref{fig:rbmgh_pes_cut0}, is further visible and clearly shows the strong role played even at short range by the electrostatic
forces. The overall anisotropy of the ground-state complex configuration is seen to be rather strong, with a sizable barrier existing for the complex bending motion outside a narrow angular cone around $\theta=0^o$. In the figure the arrow marks the minimum energy of that angular cone.
In other words, the lower charge-exchange surface is dominated by strong attractive forces on the side of the molecule to which the dipole is pointing and is locking the incoming atom via the coupling term given
asymptotically by equation~(\ref{eq:lr2}).
\clearpage
The two panels of figure~\ref{fig:rbmgh_pes_e2} further show the next higher electronic states: the lower panel reports the fitting of the raw points which describe the upper charge-exchange channel, i.e. the one
which asymptotically yields the electronically excited state of the neutral molecular partner, the \AMgH\, configuration coupled with the ionic \Rbion. It is located at about $1.6 \cdot 10^4\,cm^{-1}$ above the lowest PES of figure~\ref{fig:rbmgh_pes_e1}
and we see that essentially it maintains its strong anisotropy with a linear complex formation of minimum energy as Rb$^+$ attaches to the H-atom of the neutral molecule.
The next higher PES is the one associated with the other component of the complex in the $\Pi$ state i.e. it describes the charge-exchange compound of $A'$ symmetry. Its nearly-spherical  nature, therefore, is confirmed by the
overall PES being very flat as the angle $\theta$ varies and without the marked barrier away from linearity shown by the more asymmetrical complex on the lower panel of figure~\ref{fig:rbmgh_pes_e2}. Both components indicate
very weak interaction on the Mg-side approach of the impinging atom.
The overall anisotropy, however, changes very markedly as we look at the entrance channel side of the potential energy surface, i.e. at the electronic state of the complex which describes the $^2\Sigma^+$ state whereby the asymptotic partners are
given by the MgH$^+$ molecular ion in the X$^1\Sigma^+$ state and the neutral \Rbground\, atomic partner. The change in anisotropy is shown pictorially by the data presented in figure~\ref{fig:rbmgh_pes_e3}, where the
spherical dipole polarizability of the atomic partner drives the asymptotic behavior of the whole PES. The anisotropy of the surface presented in figure~\ref{fig:rbmgh_pes_e3} is indeed showing a smaller attractive
well located further out from the molecular center-of-mass and a more uniformly attractive region over the whole angular range.
The above findings therefore suggest a possible dynamical scheme that could be active during the initial interaction between the Coulomb crystal arrangement of the molecular ions and the ultracold Rb gas interacting with them.
In other words, the uniformly attractive features of the entrance channel PES indicate long-range attraction between partners which is largely independent of their relative orientational approach and which could thus drive the system into its geometries of stronger interactions.
A possible, pictorial path of the process is outlined over the data of figure~\ref{fig:rbmgh_pes_cut90}, where we show superimposed over the potential curves: the entrance channel access (dashed arrow on the upper right) to the region of nonadiabatic coupling between
the $^2\Sigma^+$ electronic state and the upper charge-exchange channel which produces MgH$^*$+Rb$^+$. The likely dominance of collinear orientations in that channel, and the dramatic increase of the attractive well depth in both $\Pi$
components, could drive the system to reach one of the many bound states supported by the $A'$ components of the $^2\Pi$ electronic configuration, as represented by the thick circular area. Those states could then undergo radiative stabilization into the vibrationally bound complexes of the electronic ground
state (this process is represented by the dotted downward arrow), thereby forming the neutral molecule: one can argue that the latter can then be replaced by the cold atomic ions within the Coulomb crystal environment.
To confirm the above mechanism obviously requires a more detailed analysis of its main ingredients, i.e. of the bound states supported by the [MgHRb]$^+$ complex in both the relevant electronic states after charge-exchange transitions, the spatial features and the changing strength of the corresponding transition moments, together with the actual calculation of the relevant Franck-Condon integrals. All of the above shall be the subject of an additional study, currently under way in
our laboratory, while the present work chiefly intends to extract first and discuss the specific properties of accurate ab initio potential energy surfaces which could be
employed to drive the ultracold dynamics. The general characteristics of such interactions, to our knowledge, have never been obtained before.

\section{Present conclusions}\label{sec:concl}
In this work we have computed from ab initio methods a set of potential energy curves which are expected to be involved in the low-energy interaction of a molecular polar
cation, the MgH$^+$($^1\Sigma^+$), and the neutral \Rbground\, atomic partner. The experimental setup which is considered relevant for the present study is the one provided, at ultralow
temperatures, by a Coulomb crystal arrangement of molecular ions interacting, and sympathetically cooled, by Rb atoms in a cold trap environment~\cite{schiller_2006jpb, schiller_privat}.
The analysis of the accurate calculations which have been carried out here has provided the analytic fitting of 2D grids of raw ab initio points representing the relative
strength and the orientational anisotropy of the overall interactions between the four possible asymptotic channel partners: MgH$^+$ and Rb and the three low-lying electronic
states of MgH in interaction with Rb$^+$. The results indicate the existence of  marked energy gain on charge exchange, as obviously expected from the very
beginning, but also allow us to uncover the marked orientational effects of the dynamical coupling between partners along the entrance channels and the expectedly strong
nonadiabatic coupling which would allow for transitions into  intermediate charge-exchange states that include MgH$^*$ as a partner within the complex.
The present calculations also indicate the possible radiative decay within the trap onto the ground state complex whereby the molecular ion within the Coulomb crystal can
be replaced by cold atomic ions: the latter partners are, in fact, formed within a fairly attractive potential supporting a large number of bound
states that could then break up independently of their initial geometry of formation into the intermediate state. In conclusion, the present study constitutes
the necessary, preliminary stage for
any dynamical analysis of the collisional processes occurring in the cold environment of a Coulomb crystal arrangement: it provides detailed information, which to our
knowledge has not been available before, on the spatial extension and orientational features of four of the relevant interaction potentials which can be involved in
understanding the nanoscopic dynamics of trap collisions with ultracold ions. The outcomes of the calculations locate spatial regions where nonadiabatic couplings
are more likely to play an important role, suggest possible intermediate complex formations during charge-exchange processes and indicate a possible radiative pathway
 to ionic replacement within the Coulomb crystal environment. Although no experiments as yet exist on the present complex, the information which has been gathered by
the present calculations should be a useful starting point for deciding on the likelihood of this molecular ion to be employed for sympathetic cooling with Rb cold
atomic condensates.

\section{Acknowledgments}
The financial support of the University of Rome Research Committee, of the MIUR national research project PRIN 2006 and the computational aid from the CASPUR
Consortium are gratefully acknowledged. We also thank Professor S. Schiller for suggesting this problem  to us.

\end{document}